\documentstyle[12pt]{article}
\advance\textheight by \topskip
\begin{document}

\rightline{RU-95-87}
\rightline{November 95}
\rightline{hep-th/ 9511194}

\vspace{.8cm}
\begin{center}
{\large\bf Brane -Antibrane Forces}

\vskip .9 cm

{\bf  Tom Banks}\footnote{E-mail address: banks@physics.rutgers.edu}

Department of Physics and Astronomy, Rutgers University,
Piscataway, NJ 08855-0849
\vskip.6cm

{ \bf Leonard Susskind}\footnote{E-mail address:
susskind@dormouse.stanford.edu}

Physics Department, Stanford University,
Stanford    CA 94305
\end{center}
\vskip .6 cm
\centerline{\bf ABSTRACT}
\vspace{-0.7cm}
\begin{quotation}

The force between like sign  BPS saturated objects generally
vanishes. This is a reflection of the fact that BPS states are really
massless uncharged particles  with nonvanishing momenta in
compactified directions.  Two like sign BPS objects with zero
relative velocity can be viewed as a  boosted state of two neutral
massless particles in  a state of  vanishing relative motion. By
contrast two unlike sign BPS particles  may be thought of as
colliding  objects moving in opposite directions  in compact space.
This leads to complicated  interactions  which are totally
intractable  at present.  We illustrate this by considering the
potential between opposite sign zero-D- branes .

\end{quotation}

\normalsize
\newpage

\section{Introduction }

 Perturbative string theory as presently formulated is  completely
inadequate for the study of planckian and transplanckian  processes.
No matter how small the dimensionless string coupling  $\lambda$ is,
the perturbation  expansion  breaks down at the planck
energy.  An apparent exception to this rule is presented by
BPS saturated states ,  some of
whose properties are  exactly known. Although this is true, there is
a certain sense in which  BPS physics is  often really just a
reflection  of zero energy physics.  In particular, consider those
objects which may (possibly after some duality transformation) be
thought of as Kaluza Klein charges of minimal mass. The charge
of these objects is just their momenta in the compact
dimensions.  Consider two  such BPS particles, each at rest in a
common rest frame. Suppose  that their  charges  are all proportional
to their masses.  In this case their momentum vectors (including the
compact components) are parallel. They can be thought of as massless
neutral particles with vanishing center of mass energy. This is the
class of BPS configurations for which the forces exactly cancel.
A similar pair of unlike charges correspond to a pair of oppositely
moving particles which collide with center of mass energy equal to

\begin{equation}\label{cm}
E_{cm}=2{|Q_1}{Q_2|}
\end{equation}

\noindent
where $Q_i$ is  the charge of particle $i$. These particles are
expected to undergo violent interactions, perhaps leading to black
hole formation, as the charges become large.
In this paper we consider the BPS particle states (zero branes) that
have been described by Polchinski\footnote{
Polchinski's work was motivated by earlier work of Green \cite{green}
who first made the crucial observation that Dirichlet states in the
superstring preserve half the supersymmetries.}
 \cite{joe} as D-branes.  We show that
the interactions between oppositely charged D-branes are quite different
from those between BPS states.  In particular, although one might have
thought that one could describe the dynamics by a Born Oppenheimer
potential until momentum transfers become of order the zero brane mass,
we show that the potential becomes complex and the Born-Oppenheimer
approximation breaks down when the distance between the branes is
of order the string scale.

\section{Forces Between Branes }

We consider  weakly coupled ten dimensional type 2a strings. This
theory admits zero-branes which carry Ramond-Ramond charge. These
charges can be thought of as Kaluza Klein charges in the 11
dimensional M-theory which reduces to type 2a under compactification
on a circle. Therefore the remarks of the introduction apply to these
objects. In particular two like sign branes at rest should not
interact.  In fact Polchinski has calculated the force between such
particles and finds that it vanishes. Let us briefly review the
argument.

The mass of a zero-brane is of  order $1\over \lambda$ where $\lambda$
is the type 2a
string coupling constant. Therefore when $\lambda\rightarrow 0$
the mass tends to
infinity and the particles can be treated nonrelativistically. The
calculation of the force between two like sign particles separated by
distance $Y$ is computed in lowest order as an open string  one loop
annulus diagram with the boundary conditions that the two ends of the
string are located at the sites of the two charges. Polchinski finds
the result that the potential is proportional to

\begin{eqnarray}\label{potential}
A&=& \int \frac{dt}{t}\, (2\pi t)^{-1/2}
e^{- tY^2/8\pi^2 \alpha'^2} \prod_{n=1}^\infty (1-q^{2n})^{-8}
\nonumber\\
&& \quad \frac{1}{2} \left\{ -16 \prod_{n=1}^\infty (1+q^{2n})^{8}
+ q^{-1} \prod_{n=1}^\infty (1+q^{2n-1})^{8}
- q^{-1} \prod_{n=1}^\infty (1-q^{2n-1})^{8}  \right\}
\end{eqnarray}

where $q=e^{-t/4\alpha'}$  and the integration variable $t$ is the
proper time in the open string channel. The first two terms in the
large curly bracket correspond to the exchange of NS-NS closed
strings and the third term to R-R closed strings. As pointed out by
Polchinski, the NS-NS terms cancel the R-R terms by the "usual
abstruse identity".

Now consider the effect of replacing one of the charges by an
opposite sign charge. The only effect is to change the sign of the
R-R term. The cancellation no longer occurs and the potential is
proportional to

\begin{equation}\label{unlike}
A =  \int \frac{dt}{t}\, (2\pi t)^{-3/2}
e^{-{t \over 4\alpha'}({Y^2 \over 2\pi^2}-1)} f(t)
\end{equation}
where $f(t)$ is a function which tends to $1$ for large $t$ and
rapidly tends to $0$ as $t\to 0$.

{}From this equation we first of all see that the potential
no longer vanishes. More interesting is the fact that the
integral diverges \cite{green} for
$ Y^2 < 2\pi^2 $.
To see the nature of the singularity we can
differentiate with respect to $Y$ in order to remove the dependence
on the behavior of $f$ near $t=0$. The resulting integral defines the
force and is given proportional to

\begin{equation}\label{force}
F=Y\int \frac{dt}{t}\, (2\pi t)^{-1/2}
e^{-{t \over 4\alpha'}({Y^2 \over 2\pi^2}-1)} f(t)
\end{equation}

Let  us define $Z={1 \over 4\alpha'}\left\{ {Y^2 \over 2\pi^2}-1
\right\}$ and rescale the integral to get

\begin{equation}\label{fors}
F={1 \over \sqrt{Z} }Y\int{{du \over u^{1 \over 2}}}e^{-u}f({u \over
Z})
\end{equation}
For $Z\to 0$ the function $f$ tends to $1$ and the integral becomes

\begin{equation}\label{foss}
F={1 \over \sqrt{Z} }Y\int{{du \over u^{1 \over 2}}}e^{-u} \propto{1
\over \sqrt{Z} }Y
\end{equation}

Thus we see that the attractive force diverges as $Z\to0$ and its
analytic continuation to $Y^2 <2\pi^2 $ is complex. The potential
itself is finite at $Z=0$  but is also complex for $Y^2 <2\pi^2 $.

\section{Physical Interpretation }

 A potential becoming complex is an indication that inelastic
channels are opening up. This is not too surprising since as the
opposite sign charges approach  they can annihilate into fundamental
strings. It is revealing that  the inelasticity sets in at a distance
of order $\sqrt{\alpha'}$. This seems to support the view that
D-branes carry a string scale halo around them, even though they appear
pointlike.

To get further insight  we note that the quantity  $Z$ is
proportional to the squared energy of the lightest open string
connecting the two oppositely charged zero-branes. When $Z$ becomes
negative a tachyon develops. The situation is very similar to the
thermal tachyon which occurs at the hagedorn temperature. In that
case a condensate of winding states forms when the periodic imaginary
time becomes too small. In the present case the tachyon forms when
the two charges get too close.  To follow the system beyond this
point we must compute higher order terms in the tachyon effective
potential. These can be computed in tree  approximation by
calculating multi tachyon vertex  operators. It is easy to see that
the effective potential must only contain terms even in the tachyon
field $T$. An open string can be characterized by which charges its
ends are attached to. If we call the positive charge $A$ and the
negative charge $B$  then we can classify the open strings as $AA$,
$BB$, $AB$, and $BA$. Any transition conserves the number of  $A$
ends modulo 2. Likewise for $B$ ends. We are interested in the
potential governing the $AB$ and $BA$ strings. The ground state
string of lowest energy will actually be a symmetrized sum of these
two configurations. The discrete symmetries can be seen to imply a
$Z_2$ symmetry for the effective potential for this ground state
string. The quadratic term is just the mass term proportional to $Z$.
The next term is quartic and of order $g^2$, where $g =\sqrt{\lambda}$
is the open string coupling. Thus the potential has
the form

\begin{equation}\label{potn}
V=ZT^2 +  Cg^2T^4
\end{equation}
where  the constant  $C$ is computed from the four point tachyon scattering
amplitude.

When $Z$ becomes negative the potential becomes negative and the
field rolls to its minimum at $T\sim 1/g$. This implies a condensate
containing about $1/g^2$ open strings.  If the constant $C$ is positive then
the transition is second order and the tachyon field is continuous. In
fact we find that $C$ is negative. This implies that as $Z$ decreases a
region of
metastability occurs and that the condensate is discontinuous.
We can not systematically
follow the system into this region but presumably the final state of
the evolution is the annihilation of the pair into ordinary closed
strings.  Indeed, the process should be quite similar to the
annihilation of baryon and antibaryon in large N QCD. The description of
condensate formation is also temptingly similar to the
formation of a thermal condensate near the horizon of an ordinary
black hole. It is likely that the phenomenon that we are seeing is
just  the onset  or precursor  of  non-extreme black hole formation.

At higher energies, black holes have a  large degeneracy that we
identify
with  their thermodynamic entropy. One may wonder if any sign of
degeneracy can be
found in the brane-pair system. There is one obvious source.  Due to the
symmetry  with respect to changing the sign of $T$ the condensate  can form
with either positive or negative sign. This degeneracy should be lifted by
tunnelling but the tunnelling rate is exponentially small as $g \to 0$.
Thus the system has two, essentially degenerate, states for weak coupling.
In other words, for all practical purposes,
 as we allow the branes to adiabatically approach one another, an
instability forms and the condensate can fall either way. This appears
to be a source of macroscopic irreversibility.

We suspect that the phenomenon we have discovered may be continuously
connected to black hole formation processes and that the single bit of
entropy in the two brane system will evolve into the Bekenstein-Hawking
entropy as the mass of the branes increases.  We could try to test this
by studying a system of $M$ coincident branes and $N$ coincident anti-branes
separated by a distance $Y$, in the limit that $M$ and $N$ get large.
The open string tachyon field then becomes
an $M$ X $N$ matrix and one can imagine that there will be a large
entropy associated with different stable condensates.  However, we
believe
that the single closed string exchange graph on which our calculation is
based does not contain the nonlinear gravitational physics which surely
becomes important for large mass.  Thus although the process we have
studied may indeed be continuously connected to black hole formation,
our approximation is likely to break down before we can test this conjecture.

\newpage
\section*{Acknowledgements}
The  authors would like to thank Ed   Witten for crucial remarks. One

of us (L.S.) is grateful to the Institute for  Advanced Studies and

Rutgers University for hospitality during the time this work was

being done.

\end{document}